\documentclass[twocolumn,aps,prc,superscriptaddress,showpacs,secnumroman,showkeys]{revtex4}
\usepackage{bm}
\usepackage{graphicx}
\begin{document}
\title{ The Isospin Dependence Of  The Nuclear Equation Of State Near The Critical Point} 
\author{M. Huang}
\affiliation{Cyclotron Institute, Texas A$\&$M University, 
College Station, Texas 77843}
\affiliation{Institute of Modern Physics, Chinese Academy of Sciences, 
Lanzhou, 730000,China.}
\affiliation{Graduate University of Chinese Academy of Sciences,
 Beijing, 100049, China.}
\author{A. Bonasera}
\email[E-mail at:]{bonasera@lns.infn.it}
\affiliation{Cyclotron Institute, Texas A$\&$M University, 
College Station, Texas 77843}
\affiliation{Laboratori Nazionali del Sud, INFN,via Santa Sofia, 
62, 95123 Catania, Italy}
\author{Z. Chen}
\affiliation{Cyclotron Institute, Texas A$\&$M University, 
College Station, Texas 77843}
\affiliation{Institute of Modern Physics, Chinese Academy of Sciences, 
Lanzhou, 730000,China.}
\author{R. Wada}
\email[E-mail at:]{wada@comp.tamu.edu}
\affiliation{Cyclotron Institute, Texas A$\&$M University, 
College Station, Texas 77843}
\author{K. Hagel}
\affiliation{Cyclotron Institute, Texas A$\&$M University, 
College Station, Texas 77843}
\author{J.B. Natowitz}
\affiliation{Cyclotron Institute, Texas A$\&$M University, 
College Station, Texas 77843}
\author{P.K. Sahu}
\affiliation{Cyclotron Institute, Texas A$\&$M University, 
College Station, Texas 77843}
\author{L. Qin}
\affiliation{Cyclotron Institute, Texas A$\&$M University, 
College Station, Texas 77843}
\author{T. Keutgen}
\affiliation{FNRS and IPN, Universit\'e Catholique de Louvain, B-1348 
Louvain-Neuve, Belgium.}
\author{S. Kowalski}
\affiliation{Institute of Physics, Silesia University, Katowice, Poland.}
\author{T. Materna}
\affiliation{Cyclotron Institute, Texas A$\&$M University, 
College Station, Texas 77843}
\author{J. Wang}
\affiliation{Institute of Modern Physics, Chinese Academy of Sciences, 
Lanzhou, 730000,China.}
\author{M.Barbui}
\affiliation{Cyclotron Institute, Texas A$\&$M University, 
College Station, Texas 77843}
\author{C.Bottosso}
\affiliation{Cyclotron Institute, Texas A$\&$M University, 
College Station, Texas 77843}
\author{M.R.D.Rodrigues}
\affiliation{Cyclotron Institute, Texas A$\&$M University, 
College Station, Texas 77843}

\begin{abstract}
  We discuss experimental evidence for a nuclear phase transition driven by the different concentration of neutrons to protons.
Different ratios of the neutron to proton concentrations lead to different critical points 
for the phase transition.
  This is analogous to the phase transitions 
occurring in $^{4}$He-$^{3}$He liquid mixtures. 
 We present experimental 
results which reveal the N/A (or Z/A) dependence of the phase transition and 
discuss possible implications of these observations in terms of the 
Landau Free Energy description of critical phenomena.
\end{abstract}


\maketitle

Nuclei are quantum Fermi systems which exhibit ma-\\ny interesting features which depend on temperature and density. At zero temperature and ground state density, nuclei are charged quantum drops, i.e. they have a Fermi motion~\cite{pres}  due to their quantum nature, and nucleons interact through a short range attractive
force and the long range Coulomb repulsions among the constituent protons. In the absence of the Coulomb force, the Nuclear Hamiltonian is perfectly symmetric
for exchange of protons and neutrons  apart from a small but NOT insignificant difference between the proton and neutron masses.
 This symmetry is revealed by similar energy levels in mirror nuclei, i.e. nuclei with the same mass number A but opposite numbers of neutrons, N, and protons, Z.  
 Of course this feature is observed for relatively small systems because the Coulomb energy is small~\cite{pres}. 
Analogous to the properties of  mirror nuclei, we could expect that if we study nuclei at finite temperatures, T, and low densities, $\rho$, then, if the Coulomb force is not important, the invariance under exchange
of protons to neutrons might lead to important and interesting consequences.  In fact, since the fundamental Hamiltonian of nuclei is invariant under exchange of N with Z (apart from Coulomb effects), we could expect that such an invariance should be manifested only
at high T (disordered state), while there is a spontaneous symmetry breaking at lower T (ordered state).  That means that in symmetric nuclear matter at high T,  
the state with  fragments having $N=Z$ defines the minimum of the free energy, i.e. symmetric fragments such as  deuterons and alphas would be favored at low density ~\cite{huang,land}.  On the other hand, there could be
a symmetry breaking favoring $N\ne Z$ at lower T. In this case fragments near a (first order) phase transition might prefer either a neutron or proton rich configuration. There might even be a more interesting situation, suggested by
the present data, the existence of a line of first order phase transitions ~\cite{huang} which terminates in a tri-critical point. For such a line the free energy has three equal minima: one with N=Z and the
other two for $N\ne Z$.
 Thus a phase transition is driven by the difference in isospin concentration  of the fragments $m=(N-Z)/A$.  In this paper we will discuss data which clearly demonstrate that m is an order parameter
of the phase transition.  Its conjugate field~\cite{huang} which we indicate with H, is due to the chemical potential difference
 between protons and neutrons of the emitting source at the density and temperature reached during a collision between heavy ions~\cite{prl,huang10_1}.
We also note that the phase transition has a 
strong resemblance to that observed in superfluid mixtures of 
liquid $^{4}$He-$^{3}$He near the $\lambda$ point.  In both systems, 
changing the concentration of one of the components of the mixture, 
changes the characteristics of the Equation Of State (EOS)~\cite{huang,land}.\\

In recent times a large body of experimental evidence has been interpreted 
as  demonstrating the  occurrence of a phase transition in finite nuclei 
at temperatures  (T)  of the order of  6 MeV and at densities, $\rho$, 
less than half of the normal ground state nuclear density~\cite{wci}.  
Even though strong signals for a first and a second-order phase 
transition have been found~\cite{wci,bon00}, there remain a number of 
open questions regarding the Equation of State of nuclear matter near 
the critical point.  In particular the roles of Coulomb, symmetry, 
pairing and shell effects have yet to be clearly delineated.\\

Theoretical modeling indicates that a nucleus excited in a collision expands nearly adiabatically until it is 
close to the instability region  thus the expansion is 
isentropic~\cite{siem}.  At the last stage of the expansion the role of 
the Coulomb force becomes very important.  In fact, without the Coulomb 
force, the system would require a much larger initial compression and/or 
temperature in order to enter the instability region and fragment. The 
Coulomb force acts as an external piston, giving the system an 
`extra push' to finally fragment. These features are clearly seen in 
Classical Molecular Dynamics (CMD) simulations of expanding drops 
with and without a Coulomb field~\cite{belk95,dorso}.  The expansion 
with the Coulomb force included is very slow in the later stage and nearly isothermal.\\

Even though at 
high T and small $\rho$ the nucleus behaves as a classical fluid,  the analogy to classical systems should 
not be overemphasized as, in the (T,$\rho$) region of interest,  the 
nucleus is still a strongly interacting quantum system. 
In particular the ratio of T to the Fermi energy at the (presumed) 
critical point is still smaller than 1 which suggests that the 
EOS of a nuclear system is quite different from 
the classical one. To date this expected difference has not been well 
explored~\cite{mor,wci,dago,mas,nat,ala,maria}. 

The paper is organized as follows: in the next section we discuss the experimental setup in detail.  This is followed by
a description of the data analysis and a discussion in terms of the Landau O($m^{6}$) free energy.  We, then derive
some critical exponents and the EOS corresponding to  possible scenarios suggested by our data in terms of the Fisher model of fragmentation.
Finally we draw some conclusions and suggest possible future work.

\section{Experimental Details}
The experiment was performed at the K-500 superconducting cyclotron facility at Texas
$A\&M$ University. $^{64,70}$Zn and $^{64}$Ni beams were incident on $^{ 58,64}$Ni, $^{112,124}$Sn, $^{197}$Au and $^{232}$Th
targets at 40 A MeV. Intermediate mass fragments (IMF) were detected by a detector
telescope placed at 20$^o$. The telescope consisted of four Si detectors. Each Si detector had 5cm
$\times$ 5cm area. The thicknesses were 129, 300, 1000 and 1000 $\mu$m. All Si detectors were segmented
into four sections and each quadrant had a 5$^o$ acceptance in polar and azimuthal angles.
The  fragments were detected at average angles of
  17.5$^o$ $\pm$ 2.5$^o$ and    22.5$^o$ $\pm$ 2.5$^o$. Typically 6-8 isotopes 
were clearly identified  for a given Z up to Z=18 with an energy threshold of 4-10 A MeV, using the $\Delta$E-E technique
for any two consecutive detectors. The  $\Delta$E-E spectrum was linearized by an empirical code
based on a range-energy table. In the code, isotopes are identified by a parameter
$Z_{Real}$. For the isotopes with A=2Z, $Z_{Real}$ = Z is assigned and other isotopes are identified
by interpolating between them.
The energy spectrum of each isotope was extracted by gating on lines corresponding to the individual identified isotopes. In order to compensate for the imperfectness of the linearization, actual gates for isotopes were made on the 2D plot of $Z_{Real}$ versus energy. The multiplicity of each isotope was evaluated from the extracted energy spectra, using a moving source fit at the two given angles. Since the energy spectra of some isotopes have very low statistics, the following procedure was adopted for the fits. Using a single source with a smeared source velocity around half of the beam velocity, the fit parameters were first determined from the energy spectrum summed  over all isotopes for a given Z, assuming A=2Z. Then assuming that the shape of the velocity spectrum is the same for all isotopes for a given Z. All parameters except the normalizing multiplicity parameter were assumed to be the same as for the summed spectrum.  The multiplicity for a given isotope was then derived by normalizing the standard spectrum to the observed spectrum for that isotope. 

In order to evaluate the back ground contribution to the extracted multiplicity a two Gaussian fit to each isotope peak was used with a linear background. The second Gaussian (about 10$\% $ of the height of the first one) was added to reproduce the valleys between isotopes. This component was attributed to the reactions of the isotope in the Si detector.
The centroid of the main Gaussian was set to the value calculated from the range-energy table within a small margin. The final multiplicity of an isotope with Z $>$ 2 was obtained by correction of the multiplicity evaluated form the moving source fit for the ratio between the sum of the two Gaussian yields and the linear background. 

 The yields of light charged particles (Z $\le$ 2) in coincidence with IMFs were also measured using 16 single crystal CsI(Tl) detectors of 3cm thickness  set around the target. The light output from each detector was read by a photo multiplier tube. The pulse shape discrimination method was used to identify p, d, t, h and $\alpha$ particles. The energy calibration for these particles were performed using Si detectors (50 -300 $\mu$m) in front of the CsI detectors in separate runs. The yield of each isotope was evaluated, using a moving source fit. Three sources (projectile-like(PLF), nucleon-nucleon-like(NN) and target-like (TLF)) were used. The NN-like sources have source velocities of about a half of the beam velocity. The parameters were searched globally for all 16 angles. Detailed procedures of the data analysis are also given in refs.~\cite{huang10_2,chen10}.

\begin{figure}[ht]
\centerline{
\includegraphics[width=3.0in]{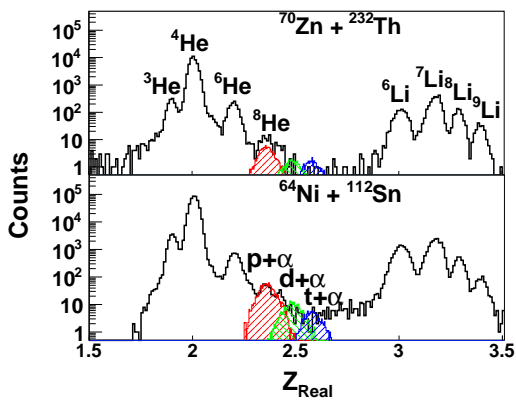}}
\caption{Typical $Z_{Real}$ spectra for He and Li isotopes. Accidental events are generated only for p, d, t +$\alpha$ and shown separately by shaded histograms as indicated.
}
\label{fig:fig_0}
\end{figure}

Special care has been taken for $^{8}$He identification. All He isotopes are identified in the Si telescope, using the $\Delta$E-E technique, in a narrow energy range. When a proton and an $\alpha$ hit the same quadrant and when both of them stop in the E detector, their $\Delta$E-E points overlap with those of $^{8}$He. Since the multiplicities of protons and alphas are about three orders of magnitude larger than that of $^{8}$He, the contribution of accidental events becomes significant, especially for the reaction systems with lower numbers of neutrons, in which $^{8}$He production is suppressed. Since Z=1 $\Delta$E-E spectra are not available in this experiment, $\Delta$E-E for Z $\le 3$ were measured in a separate run. Using the light charged particle multiplicity extracted from the 16 CsI detectors, the accidental events were simulated for each reaction for the observed $\alpha$ yield in the $\Delta$E-E spectra in this experiment, the solid angle of the quadrant and the multiplicity of Z=1 particles. In order to minimize the accidental events, the runs with a low beam intensity were selected in each reaction. Typical linearized $Z_{Real}$ spectra with these accidentals are shown in Fig.(\ref{fig:fig_0}) for $^{70}$Zn + $^{232}$Th (N/Z $\sim$ 1.5) and $^{64}$Ni + $^{112}$Sn ( N/Z $\sim$ 1.25). As one can see, the $Z_{real}$ values for the accidental events of proton and $\alpha$  pileup is nearly identical to that of $^{8}$He, while $^{6}$He is clearly identified. The contributions from d + $\alpha$ and t + $\alpha$ are also reasonably consistent with the observed background yields. A significant excess of $^{8}$He yield beyond the accidentals is only observed for the reaction systems with the $^{124}$Sn, $^{197}$Au and $^{232}$Th targets. After the correction of the accidental contributions, the multiplicities of $^{6}$He and $^{8}$He were calculated using the source fit parameters obtained for Li isotopes.    

\section{Data Analysis}
\begin{figure}[ht]
\centerline{
\includegraphics[width=3.0in]{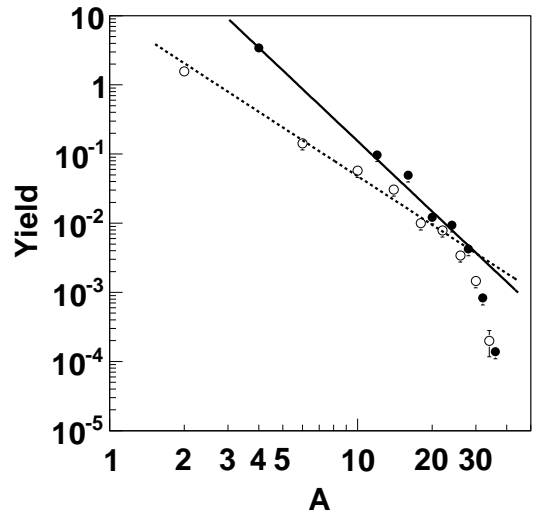}}
\caption{Mass distribution for the $^{64}$Ni+$^{124}$Sn system 
at 40 MeV/nucleon for I=0. The 
lines are power law fits with exponents $2.3\pm0.02$ (odd-odd nuclei, dashed line), and
$3.4\pm0.06$ (even-even nuclei, full line) respectively. 
}
\label{fig:fig_1}
\end{figure}

The key factor in our analysis is the value $I=N-Z$
of the detected fragments.  
A plot of the yield versus mass number when I=0 displays a power law 
behavior with yields decreasing as $A^{-\tau}$ ~\cite{min,prl}.  This is shown in Fig.(\ref{fig:fig_1}) for the 
$^{64}$Ni+$^{124}$Sn case at 40 MeV/nucleon.   In the figure  we have made separate fits for 
 odd-odd (open symbols) or even-even(filled symbols) nuclei. 
 As seen,   different exponents $\tau$ appear which suggests that pairing is 
playing a role in the dynamics ~\cite{pres}, leading to higher yields for even-even nuclei.

The observation of the power law behavior suggests that the mass distributions may 
be discussed in terms of a modified Fisher model~\cite{bon00,min}:  
\begin{equation}
  \label{eq:eq_1}
  Y=y_0A^{-\tau}e^{-\beta \Delta \mu A},
\end{equation}
where y$_{0}$ is a normalization constant, $\tau=2.3$ is a critical exponent ~\cite{bon00}, $\beta$ is the inverse 
temperature and $\Delta\mu = F(I/A)$ is 
the free energy per 
particle, $F$,  near the critical point. 
Recall that in general, the free energy is a function of the mass A (volume), $A^{2/3}$(surface) and the chemical composition $m$ of the fragments and possibly pairing.  
The region we are studying in this paper seems near the critical point for a liquid gas phase transition (volume and surface equal to zero) but modified by $m=\frac{I}{A}$.  Because of this modification we can
observe different features of the transition such as a first order phase transition driven by $m$, the order parameter.

We begin our analysis by noting that the Fisher free energy is usually written in terms of the volume and the surface of a 
drop undergoing a (second order) phase transition~\cite{fisher}. Our data indicate that those terms are not important in the present case~\cite{bon00} as we will show more in detail below.  If they are negligible this suggests that we are near
the critical point for a liquid gas phase transition.  Because we have two different interacting fluids, neutrons and protons, the transition becomes more complex and more interesting than in a single component liquid.
Experiments at different energies might display a free energy which depends on all these factors. If we accept that $F$ is dominated 
by the symmetry energy we can make the approximation that 
$F(I/A)=E_{sym}= 25(I/A)^{2}$ MeV/A,  i.e. the symmetry energy of a nucleus in 
its ground state~\cite{pres}.  We will use this relationship in order to 
infer an approximate value of the temperature of the system.  However, 
we stress that in actuality, F(I/A) is a function of density, temperature 
and all other relevant quantities near the critical point.
According to the Fisher equation given above, we can compare all systems 
on the same basis by normalizing the yields and factoring out the power 
law term.  For this purpose we have chosen to normalize the yield data for each system 
to the $^{12}$C yield ($I=0$) in that system, i.e. we define a ratio:

\begin{equation}
  \label{eq:eq_2}
  R =\frac{Y A^{\tau}}{Y(^{12}C) 12^{\tau}}.
\end{equation}

\begin{figure}[ht]
\centerline{
\includegraphics[width=3.00in]{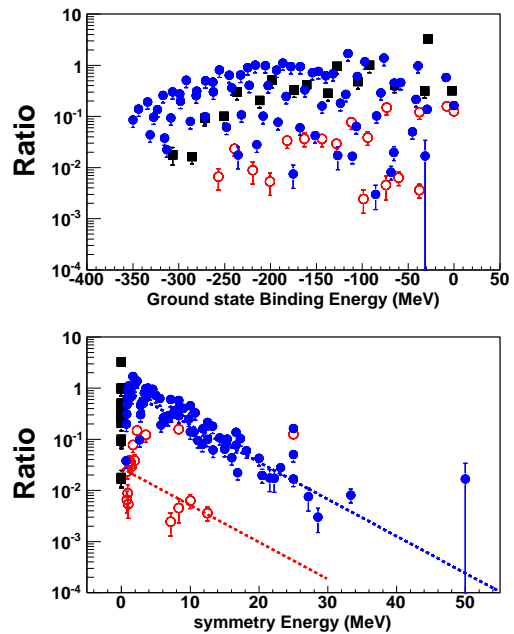}}
\caption{
Ratio versus fragments ground state binding energy (top panel) and symmetry energy (bottom panel) for the $^{64}$Ni + $^{64}$Ni case at 40 MeV/nucleon. $a_sym$=25 MeV is used. The $I < 0$ and $I > 0$ ($I=0$)isotopes are indicated by the open and full circles respectively (full squares). The dashed lines (bottom panel) are fits using a ground state symmetry energy, Eq.\ref{eq:eq_1}, and a `temperature' of 6 MeV. Notice that the given experimental $^{8}$He yield is the upper limit.
}
\label{fig:fig_2}
\end{figure}

\begin{figure}[ht]
\centerline{
\includegraphics[width=3.00in]{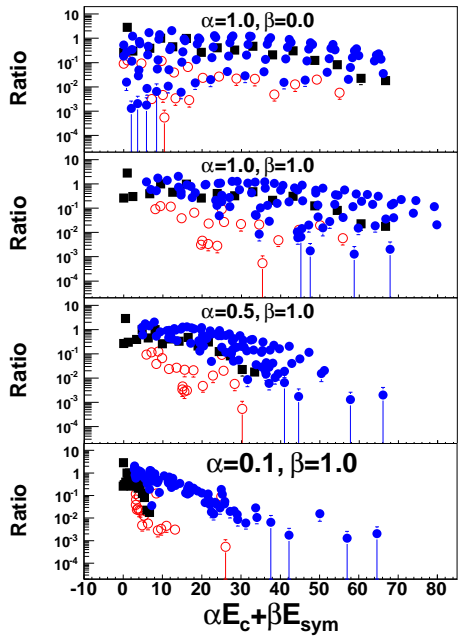}}
\caption{
Ratio versus symmetry energy + Coulomb energy for the $^{64}$Ni + $^{124}$Sn case at 40 
MeV/nucleon.  The panels from top to bottom are for different combinations of the symmetry and Coulomb energy.  The 
$I < 0$ and $I > 0$ ($I=0$)isotopes are indicated by the open and full circles 
respectively (full squares). 
}
\label{fig:fig_3}
\end{figure}

\begin{figure}[ht]
\centerline{
\includegraphics[width=2.75in]{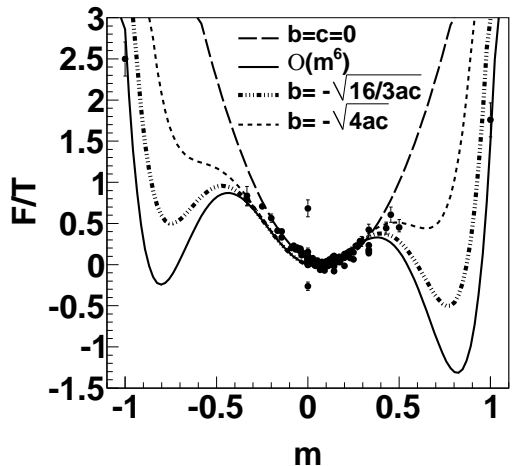}}
\caption{
Free energy versus m  for the case $^{64}$Ni+$^{232}$Th.  
The  full line is a free fit based on Landau O($m^{6}$) free energy. The dashed-dotted-dotted-dotted line is obtained imposing in the fit
$b=-\sqrt {16/3 ac}$ and it is located on a line of first order phase transitions.  The short dashed line corresponds to $b=-\sqrt {4 ac}$,
i.e. superheating. The $O(m^2)$ case, $F/T=a(m-m_{s})^2, $i.e. $b=c=0$, $m_{s}=0.1$, is given by the long dashed line.
}
\label{fig:fig_4}
\end{figure}

The normalized ratios for the system 
$^{64}$Ni + $^{64}$Ni at 40 MeV/nucleon are plotted as a function of the 
(ground state) symmetry energy in Fig.(\ref{fig:fig_2}), bottom panel.  The data display an 
exponential decrease with increasing symmetry energy, except for the 
isotopes for which $I = 0$.  The yields of these I = 0 isotopes are 
of course not sensitive to the symmetry energy but rather to the Coulomb 
and pairing energies and possibly to shell effects. 
 A fit to the exponentially decreasing portion of the data using the ground state symmetry energy  gives an `apparent 
temperature' T of 6.0 MeV.  This value of T would be the 
real one if only the symmetry energy is important, if entropy can be neglected, if $a_{sym} = 25$ MeV (the g.s. symmetry energy coefficient value) 
and if secondary decay effects are negligible.  In general we expect that 
the symmetry energy coefficient is density and temperature dependent. Further, secondary decay processes may modify the primary fragment distributions~\cite{huang10_2,chen10}.   
we will discuss these questions in the framework of the Landau free energy approach below.  
We stress that the appearance of two branches in Fig.\ref{fig:fig_2} (bottom), indicates 
that the total free energy must contain an odd power term in (I/A)  at variance 
with the common expression for the ground state symmetry energy. For reference in the top part of Fig.(\ref{fig:fig_2}) we have plotted the ratio versus the total ground state binding energy of the fragments.  No clear correlations are observed which
might suggest that the symmetry energy dominates the process. \\
It is surprising that such a scaling appears as a function of the symmetry energy  only.  In fact we might wonder about the role of the Coulomb energy if we accept that surface and volume terms give negligible contribution. 
In figure \ref{fig:fig_3} we have plotted the same normalized ratios as a function of the quantity $\alpha E_{coul}+\beta  E_{sym}$, $\alpha$ and $\beta$ are arbitrary parameters given in the figure 
and
$E_{coul}=0.7 Z(Z-1)A^{-1/3}$ is the Coulomb contribution to the ground state energy of the nucleus. We see from the figure that 
by decreasing the relative contribution of the Coulomb energy compared to the symmetry energy the scaling appears.  This implies that the Coulomb energy is much less
important than the symmetry energy near the critical point, which suggests that either the density dependence of  those two terms is different or
that, at the time of formation the fragments are strongly deformed, reducing the Coulomb effect.  Such deformations have been seen in 
CMD calculations of fragmentation ~\cite{bon00}.\\
To further explore the role of the relative nucleon concentrations we plot 
in Fig.(\ref{fig:fig_4}) the quantity $\frac{F}{T}=-\frac{ln(R)}{A}$  versus $m = (I/A)$, 
the difference in neutron and proton concentration of the fragment.  As expected the normalized yield 
ratios depend strongly on m.

Pursuing the question of phase transition 
we can perform a fit to these data within the generalized Landau free energy description~\cite{huang}. In this 
approach the ratio of the free energy to the temperature is written in 
terms of an expansion:
\begin{equation}
  \label{eq:eq_3}
 \frac{F}{T}=\frac{1}{2}a m^2+\frac{1}{4}b m^4 +\frac{1}{6}c m^6-m\frac{H}{T},
 \end{equation}
where $m$ is an order parameter, $H$ is its conjugate variable and 
$a-c$ are fitting parameters
~\cite{huang}.  We observe that the Free energy is even in the exchange of $m \rightarrow -m$, reflecting the invariance of the nuclear
forces when exchanging N and Z.  This symmetry is violated by the conjugate field $H$ which arises when the source is asymmetric
in chemical composition.  We stress that m and H are related to each other through the relation $m=-\frac{\delta F}{\delta H}$.

The use of the Landau approach is for guidance only. While the approximation 
to $O(m^4)$ does not work~\cite{prl}, the $O(m^6)$ case is in good agreement with the data.  This is not surprising since, if fluctuations are important, a higher order approximation to the free energy is better, i.e. gives
critical exponents closer to those seen in the data and satisfies the Ginzburg criterion ~\cite{huang}.
 A free  fit using Eq.\ref{eq:eq_3} 
 is displayed in 
Fig.\ref{fig:fig_4} (full line).  
Notice the change of curvature near $m=0.3$, which incidentally is close to $m_{cn}$ of the compound nucleus. 
For comparison in the same figure we have displayed the $O(m^2)$ case, i.e., $F/T=a(m-m_s)^2$ ($b=c=0$) \footnote{$F/T=a(m-m_s)^2=(a/2)m^2-H/Tm+(a/2)m_{s}^2; H/T=am_(s)$. The last term is dropped out when the yields are normalized by $^{12}$C.}
As seen in the plot last assumption also produces a reasonable fit, although it does not reproduce shoulders near m $\sim \pm$0.3. As we will discuss in more detail below the appearance of two minima for $m\ne 0$ (when $H/T=0$) might be a signature for the existence of a first order phase transition  occurring in these reactions.


In general the coefficients entering the Landau free energy  Eq.(\ref{eq:eq_3}), depend on  temperature, pressure or density
of the source.  Usually one assumes $c>0$, $a=a_0(\rho)(T-T_0)$ and $b=b(T,\rho)$, where $T_0$ is some `critical' temperature discussed below.
 The precise determination of these
parameters determines the nuclear equation of state (NEOS) near the critical point.  The data we have do not allow such a complete constraining of the 
NEOS but do suggest some interesting  possible scenarios which we discuss below. 

\begin{figure}[ht]
\centerline{
\includegraphics[width=6.3cm,clip]{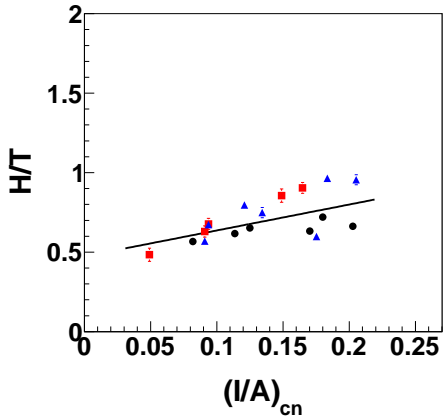}}
\caption{
H/T versus (I/A) of the compound nucleus obtained from the data fit to the Landau free energy, Eq.(\ref{eq:eq_3}). The full
circles are for  $^{64}$Ni, the full triangles are for  $^{70}$Zn and the squares for $^{64}$Zn projectiles
impinging on various targets, see text.
}
\label{fig:fig_5}
\end{figure}

\begin{figure}[ht]
\centerline{
\includegraphics[width=6.3cm,clip]{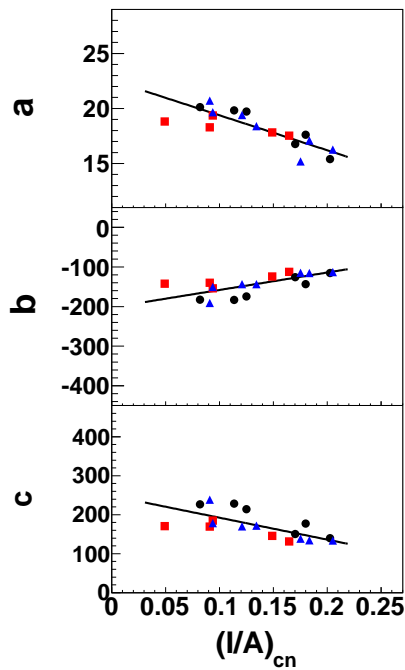}}
\caption{
The parameters a,b and c  versus (I/A) of the compound nucleus obtained from the data fit to the Landau free energy, Eq.(\ref{eq:eq_3}). The symbols are like in figure \ref{fig:fig_5}.}
\label{fig:fig_6}
\end{figure}

We begin by noting that the conjugate variable H which appears in equation (\ref{eq:eq_3}) is determined by the chemical composition  of the source.
Since, in general, the source has $N \ne Z$, the extreme of $F/T$ are displaced from the values obtained when H=0.  In fact if we take the first derivative of
the free energy we get:
\begin{equation}
  \label{eq:eq_4}
( \frac{F}{T})'=a m+b m^3 +c m^5-\frac{H}{T}.
 \end{equation}
When $H/T=0$ the first derivative is zero for the following values of m ~\cite{huang}:
\begin{equation}
  \label{eq:eq_5}
m_0=0;
m^2_{\pm}=\frac{-b \pm \sqrt {b^2-4ac}}{2c}.
 \end{equation}
If we now assume $H \ne 0$ but small, we can expand the solutions above 
as $m=m_{0\pm}+\eta$ with $\eta$ small.  Equating  the first derivative to zero, Eq.(\ref{eq:eq_4}),  and neglecting terms $ O(\eta ^2)$ we get:
\begin{equation}
  \label{eq:eq_6}
 \eta=\frac{H/T}{a+3bm_{0\pm}^2+5cm_{0\pm}^4}.
 \end{equation}
The shift of the minimum from $m_0=0$ should be given by the equation above and should be proportional to
$m$ of the emitting source.  We can easily check this feature in our data.  In Fig.(\ref{fig:fig_5})
we plot the values of $H/T$ obtained from the fits to our data for all systems using Eq.(\ref{eq:eq_3}) versus $m_{cn}=(I/A)_{cn}$.  \\
The linear fit in Fig.\ref{fig:fig_5} is given by $H/T=0.47+1.6(I/A)_{cn}$ which agrees with the linear dependence of Eq.(\ref{eq:eq_6}).
However, for this fit  $H/T\ne 0$  for $I_{cn}=0$  which could indicate the favoring of  $N>Z$ fragments by the Coulomb field. Another possibility is that $(I/A)_{source}\propto (I/A){cn}$ which then gives $H=0$ when $I_{source}=0$. 
Finally we should consider that together with H also the temperature may also be changing some since the collisions are between different target-projectile combinations at the same beam energy.  If the temperature were the same then the coefficients of
the free energy, Eq.\ref{eq:eq_3},  should be independent of the source size, only H/T should change.  In Fig.(\ref{fig:fig_6}) we plot the parameters a, b and c as a function of
the compound nucleus $m_{cn}$. As we see there is some dependence which may reflect differences in temperature.  However, we note that
the error bars and fluctuations are large which may also  indicate important secondary  decay effects. Thus is not too so easy to draw definite conclusions.

\begin{figure}[ht]
\centerline{ 
\includegraphics[width=6.3cm,clip]{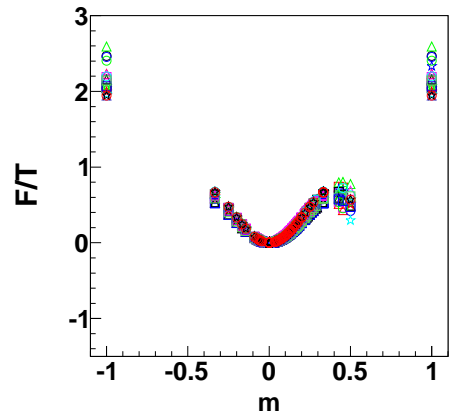}}
\caption{
F/T (H/T=0)   versus m of the fragments obtained from the a, b and c parameters fit to the Landau free energy, Eq.(3). Results for all experimentally investigated reactions are displayed.}
\label{fig:fig_7}
\end{figure}

Given the  information on the parameters of the Landau free energy contained in figures \ref{fig:fig_5} and \ref{fig:fig_6} we can discuss some features regarding the NEOS.  In particular for each reaction system we
can estimate F/T when $H/T=0$.  In figure \ref{fig:fig_7} we plot this quantity versus m of the fragments  for various reactions.  
The curves do not differ much suggesting that temperatures are quite similar.  The fits exhibit curvature near $m =\pm 0.4$ which  may suggest
 the presence of additional minima at larger absolute values of m. This could indicate either a first order phase
transition or superheating (see below). The lack
of data at very large $m$ makes it difficult to constrain the fit. However, we can study other situations of particular physical interest which arise when the relationships among the parameters a, b and c are constrained. ~\cite{huang,land,prl}:\\

We have considered four such cases as follows:

1) superheating.  This case corresponds to $b=-\sqrt {4ac}$ and gives two minima at $m\ne 0$ and is plotted in figure \ref{fig:fig_4} for the $^{64}$Ni+$^{232}$Th system with a short dashed line.
  These are not absolute minima,  which occur  only at $m=0$, and they correspond to 
metastable states.  They might be observed in high quality data for collisions of more neutron rich or proton rich systems making a hot source with
$m_{source}\approx  \pm 0.4$.  In fact if the system could be gently brought to the right temperature  $T_s$, with the correct isotopic composition, it might stay in the minimum, i.e.
more fragments of that $m$ should appear;\\
2) line of first order phase transition.  This corresponds to the condition $b=-\sqrt {16ac/3}$  at a  temperature  $T_3$, 
which, if imposed on the fit of the free energy, results in the dashed-dotted-dotted-dotted line of
Fig.\ref{fig:fig_4}. This fit is of similar quality to the previous cases. Now the minima are at $m\approx 0.6$, i.e. for more neutron rich fragments due to the fact that $H/T\ne 0$. This suggests that in this
situation we might produce a large number of neutron rich fragments.  However, most of those fragments are probably unstable, thus coincidence measurements may be required to determine
their yields.  Of course this feature should become important in neutron rich stars. \\
3) first order phase transition.  Corresponds to the case $a=0$ and determines the `critical temperature' $T_0$ where the minimum at $m=0$ disappears and only the ones at
$m\ne 0$ survive. This case is excluded by our present data. However, the fit in Fig.(\ref{fig:fig_4}) suggests an intermediate situation between this and case (2) above.\\
4) line of second order phase transition, tri-critical point.  Corresponds to $a=0$ and $b>0$ ($T=T_c$).  When $b=0$ as well we have a tricritical point ($T=T_{3c}$),
 i.e. the point where the line of first order
phase transition terminates into a second order phase transition.  This case is also excluded by our data.\\
\begin{figure}[ht]
\centerline{
\includegraphics[width=6.3cm,clip]{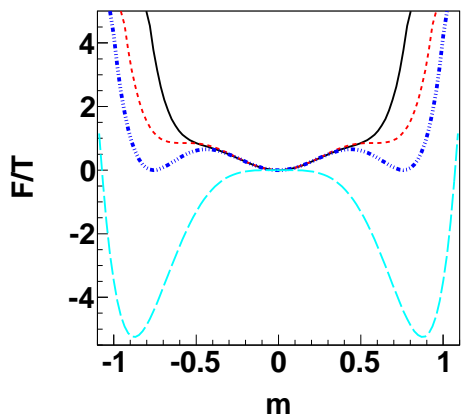}}
\caption{
F/T (H/T=0)   versus m of the fragments obtained from the a, b and c parameter fit to the Landau free energy, Eq.(\ref{eq:eq_3}) for $^{64}$Ni+$^{232}$Th.  The four curves correspond
to vapor(full line), superheating(short dashed), line of first order transition(3 critical dashed-dotted-dotted-dotted line), first order phase transition(long dashed line), see text.}
\label{fig:fig_8}
\end{figure}
We can extrapolate the cases discussed above to   $H/T=0$ as was done for Fig.\ref{fig:fig_7}.  In Fig.\ref{fig:fig_8} we plot F/T (H/T=0) (extrapolated from the data) vs. $m$. 
Purists will not call this the EOS but reserve that for the pressure vs. $m$ case (that we discuss below).  Since H/T is zero, the curves are symmetric with respect to $m$.
We see in the plot: vapor (dashed-dotted-dotted line) $T> T_s$, superheating $(T=T_s)$(dotted line),
 a point in the line of a first order phase transition $(T=T_3)$(dashed-dotted line) which displays three equal minima at $m_0$ and $m_\pm$, see Eq.\ref{eq:eq_5}, the experimental data fit line(full line)$T<T_3$. We have also added the case $a=0$ which should be obtained at $T=T_0$, where the minimum at $m=0$ becomes a maximum.  There is a series of cases not displayed in the figure, corresponding to the temperatures between $T_0$ and $T_3$ where $m=0$ is still a minimum but $\it {not}$ an absolute minimum. This corresponds to
supercooling and might be observed in gentle collisions of $N=Z$ nuclei similarly to the superheating case.

The  features in Fig.\ref{fig:fig_8} are reminiscent of the superfluid $\lambda$ transition 
observed as some $^{3}$He is added to $^{4}$He ~\cite{huang}. Pure $^{4}$He  has a critical temperature of 2.18 K. The 
critical temperature for the second-order transitions decreases with 
increasing $^{3}$He concentration until at temperature,  
$T = 0.867$K, a first-order transition appears. This point is known as 
the tri-critical point for this system. In a similar fashion, a  nucleus, which can 
undergo a liquid-gas phase transition, should be influenced by the different 
neutron to proton concentrations.  Thus the discontinuity observed in 
Fig.\ref{fig:fig_4} ($m=0$) could be a signature for a tri-critical point as in the 
$^{4}$He-$^{3}$He case.  We believe that our data, analyzed in terms of the 
the Landau O($m^{6}$) free energy, suggest such a feature but 
are not sufficient to clearly demonstrate this.  Some other 
work~\cite{campi,gulm}, also suggests that a line of critical points 
might be found away from its `canonical' position, i.e. at the end of 
a first-order phase transition and, for small systems, even extending 
into the coexistence region.  
\section{Critical exponents}
In the fits discussed above the parameters a, b and c were left free since we do not have any particular values to fix the scale.  Nevertheless, we saw in figure \ref{fig:fig_7} that the free energy (H/T=0) looks very similar for the different systems.  Thus the values of the fitting parameters are similar apart from a scaling factor.
We can  avoid unnecessary factors by defining suitable dimensionless quantities.  This can be accomplished by looking at the solutions of the minima of the free energy, cf. 
Eq.\ref{eq:eq_5}.  In particular from the value at the minimum, $m_+$ we can define the following quantities  ($b\ne 0$):
\begin{equation}
  \label{eq:eq_7}
 x=\frac{4ac}{b^2}.
\end{equation}
Recalling that $a$ is related to the distance from the critical temperature while b and c should only depend on density ~\cite{huang}, we deduce that
x is a measure of the distance $T-T_3$ from the critical temperature in a suitable dimensionless fashion.  Similarly we can define a reduced order parameter from Eq.\ref{eq:eq_5}:
\begin{equation}
  \label{eq:eq_8}
y=\frac{2c m^2}{|b| }.
\end{equation}
Thus Eq.\ref{eq:eq_5} can be rewritten as:
\begin{equation}
  \label{eq:eq_9}
 y=1+\sqrt{1-x}.
\end{equation}
Near the critical point we know that the order parameter has a singular part that behaves in a power law fashion, thus we can define the singular part as:
\begin{equation}
  \label{eq:eq_10}
M=\pm \sqrt{y-1}=\pm {(1-x)}^{1/4},
\end{equation}
 defining the temperature `distance' from the critical point, $|t|=|1-x|$, immediately gives the value of a critical exponent: $\beta=\frac{1}{4}$.  
This exponent is very close to the accepted experimental value as is well known in the
O($m^6$) Landau theory ~\cite{huang}.
\begin{figure}[ht]
\centerline{
\includegraphics[width=2.75in,clip]{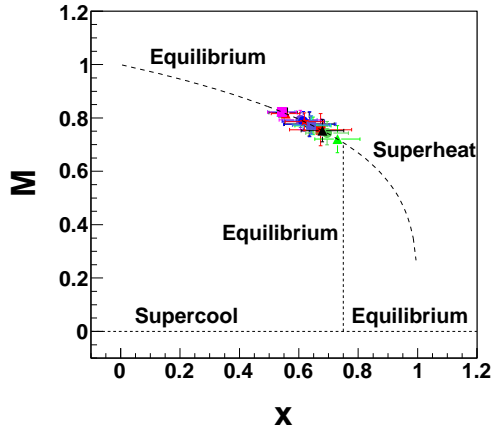}}
\caption{
Order parameter versus reduced temperature for all studied systems. The dashed line is given by Eq.(\ref{eq:eq_10}), the vertical line indicates the critical temperature $T_3$.  To the right
of this line the system is in a superheated state. Supercooling occurs on the left of the vertical line and M=0~\cite{huang}.   }
\label{fig:fig_9}
\end{figure}
In Fig.(\ref{fig:fig_9}) the experimental values of M and x obtained within the Landau  theory are plotted together with the equilibrium condition given by Eq.(\ref{eq:eq_10}).  Supercooling and
Superheating regions,  as discussed in the previous section, can be identified as well ~\cite{huang}.

\begin{figure}[ht]
\centerline{
\includegraphics[width=2.75in,clip]{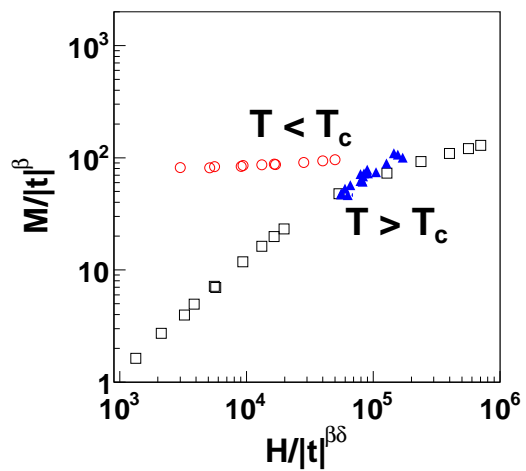}}
\caption{
Scaling form for magnetization M vs. external field for nickel~\cite{huang}, open symbols.  The corresponding quantities for nuclei normalized to the metal case are given by the full symbols. }
\label{fig:fig_10}
\end{figure}

As is the case for macroscopic systems we can now `turn ' the external field H on and off. In our case this is done with a suitable choice of the colliding systems.  In this way we can study the `EOS'
at the critical point by turning on H:
\begin{equation}
  \label{eq:eq_11}
M=H^{1/\delta},
\end{equation}
which defines the critical exponent $\delta$. In the Landau theory this exponent can be determined at the critical point where $a=b=0$. From equation (\ref{eq:eq_4}) we easily get  $\delta=5$, which is the accepted value for such a critical exponent~\cite{huang}.
 In order to exactly determine this exponent we need to bring the system to the critical point. This does not appear to be the case for our data
as we saw in figure \ref{fig:fig_9}.  Nevertheless a plot of the order parameter versus H should display a power law behavior as it is well known in macroscopic systems~\cite{huang}.
A precise determination of the critical exponent requires the knowledge of the temperature T both above and  below 
the critical point.  This is feasible but requires precise experimental data.  
>From Eq.(\ref{eq:eq_6}) assuming the only minimum is at m=0 we get
\begin{equation}
  \label{eq:eq_12}
\eta=\frac{H/T}{a}.
\end{equation}
The temperature (a) dependence of the order parameter shows that we are away from the critical point.  Nevertheless we can study the behavior  close to
the critical point by suitably defining scaling forms ~\cite{huang}: $\frac{M}{|t|^\beta}=\frac{\eta}{|t|^\beta}$ vs. $\frac{H/T}{|t|^{\beta \delta}}$.  These quantities are plotted in
figure \ref{fig:fig_10} and compared to magnetization data for nickel metal. The scaled magnetization is plotted versus the scaled external magnetic field~\cite{huang}.  The nuclear data have been shifted in the region near
the crossing of data above and below the critical temperature where we expect our data to be, see Fig.\ref{fig:fig_9}.  Of course it is not possible at this stage to directly compare to the macroscopic data since we have
no information for the absolute values of the temperatures.  Furthermore   the role of the density (or pressure) is not clear since we expect that the parameter $a$ (or equivalently x) depends on the `distance' from
the critical temperature and critical pressure. These quantities could however be obtained in $4\pi$ experiments where charges, masses and their velocities are carefully determined.   

Once we have derived the `reduced' parameters of the Landau $O(m^6)$ theory, we can write a `reduced' free energy as ($b\ne 0$):
\begin{equation}
  \label{eq:eq_13}
\frac{f}{T}=\frac{1}{2}xm^2-|z|m^4+\frac{2}{3}z^2m^6-\frac{{\it h}}{T} m,
\end{equation}
where: $\frac{f}{T}=\frac{4c}{b^2}\frac{F}{T}$, $z=\frac{c}{|b|}$, $\frac{h}{T}=\frac{4c}{b^2}\frac{H}{T}$.  These quantities together with the temperature, Eq.(\ref{eq:eq_7}),
and the reduced order parameter y,of Eq.(\ref{eq:eq_8}), constitute the Landau $O(m^6)$ theory in dimensionless form.  It is instructive to study how these quantities change with the reaction system as
we did in figures (\ref{fig:fig_6}) and (\ref{fig:fig_7}).  
\begin{figure}[ht]
\centerline{
\includegraphics[width=6.3cm,clip]{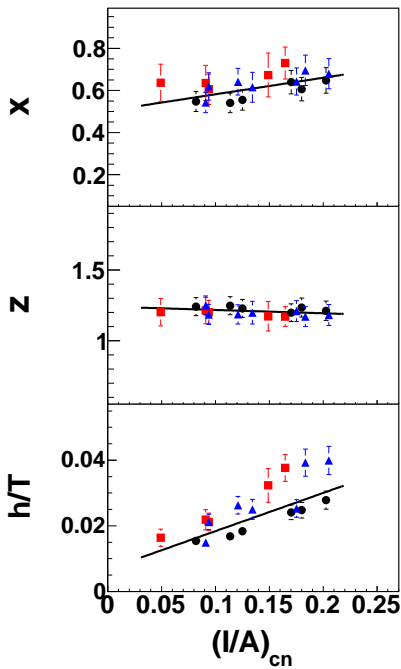}}
\caption{
The parameters x, z and h/T  versus (I/A) of the compound nucleus obtained from the data fit to the Landau free energy, Eq.(\ref{eq:eq_3}). The symbols are like in figure \ref{fig:fig_5}.}
\label{fig:fig_11}
\end{figure}
In Fig.(\ref{fig:fig_11}) we plot these normalized quantities vs. difference in neutron proton concentration of the compound nucleus. Compare to figure \ref{fig:fig_6}.  A feature worth noticing is the following, while the parameter $a$ 
is decreasing with increasing  (I/A) of the compound nucleus, the opposite holds for the parameter $x$ which gives the `distance' from the critical temperature, see Fig.(\ref{fig:fig_11}). This is very important since
only normalized quantities should be used when inferring the properties of the EOS (i.e. temperature, density etc.) near the critical point.
\section{Symmetry and Pairing compared to the Coulomb Energy.}
\begin{figure}[ht]
\centerline{
\includegraphics[width=6.3cm,clip]{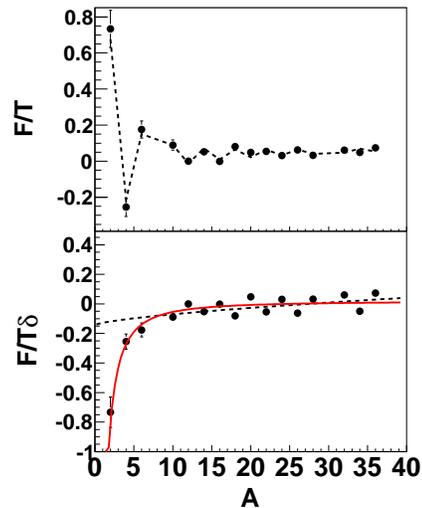}}
\caption{
Free energy versus mass for $m=0$ isotopes for the  $^{70}$Zn+$^{124}$Sn system (top panel), the dashed line is a fit using Coulomb and pairing contributions. 
Free energy times $\delta$ (see text-bottom panel) versus mass for $m=0$ isotopes. The lines are separate fits suggested by the Coulomb(dashed line) and pairing(full line) energy mass number dependence.}
\label{fig:fig_12}
\end{figure}

In the previous sections we have seen that the Coulomb energy might become important especially for large values of the charges.  We can now try to derive some
qualitative understanding of  when and why Coulomb corrections might become important and might even hinder a possible phase transition.  From the mass formula we can write the Coulomb energy for large Z as~\cite{pres}:
\begin{equation}
  \label{eq:eq_14}
 \frac{E_c}{A}=0.77 \frac{Z^2}{A^2}A^{2/3}=\frac{0.77}{4}(1-m)^2 A^{2/3},
\end{equation}
which explicitly introduces the order parameter $m$ in the Coulomb energy.   We can define an `effective' symmetry energy (per particle) as:
\begin{equation}
  \label{eq:eq_15}
 \frac{E_{eff}}{A}=
 (a_{sym}+\frac{0.77}{4}A^{2/3})m^2-\frac{0.77}{2}A^{2/3}m+\frac{0.77}{4}A^{2/3},
\end{equation}
where the symmetry energy coefficient $a_{sym}=25 MeV$.  Ignoring for a moment density corrections we see that $O(m^2)$ term should be affected by Coulomb corrections for large fragment mass numbers.  Furthermore, a linear term in $m$ is introduced which will then modify the `external' field even in collisions where the source $m_s=0$ as we discussed in Fig.(\ref{fig:fig_5}).
Finally there is a term not dependent on $m$ that will destroy the scaling for large mass (charge) numbers.  We should also notice that assuming a spherical expansion, at low densities the Coulomb energy will decrease as $\rho^{1/3}$ 
while contributions to the symmetry energy should depend both on  $\rho^{2/3}$ reflecting the  Fermi energy of the nuclei and on $\rho$, the latter coming from different n-p interactions. At low densities we
would expect Coulomb to be stronger  than it appears to be in the data.  This may be indicative that the fragments must be highly deformed, reducing the Coulomb energy.
Coulomb corrections should become more important when $m=0$ for the detected fragment.  We have plotted the yields of $m =0$ nuclei in Fig.(\ref{fig:fig_1}) and pointed out that pairing appears to be playing a role.  From Eq.(\ref{eq:eq_15}) above we
should expect that, if Coulomb is dominant for such fragments, the free energy should depend on $A^{2/3}$.  In Fig.(\ref{fig:fig_12}) (top panel) we plot F/T versus $A$ for $m=0$ fragments.  
The expected dependence with mass number in the free energy suggested from `effective' symmetry energy Eq.(\ref{eq:eq_15}) is not seen in the figure.
Rather, a staggering between odd-odd and even-even nuclei is clearly visible.  

To better clarify these arguments we can write the pairing energy from the mass formula as~\cite{pres}:
\begin{equation}
  \label{eq:eq_16}
 \frac{E_p}{A}=12 \frac{\delta}{A^{3/2}},
\end{equation}
where $12 MeV$ is the ground state pairing energy coefficient and $\delta$:
\begin{equation}
  \label{eq:eq_17}
 \delta= \frac{(-1)^N+(-1)^Z}{2}.
\end{equation} 
The suggested mass dependence from pairing, Eq.\ref{eq:eq_16}, is completely different from the Coulomb one when $m=0$, see Eq.\ref{eq:eq_14}.

Notice that it is the $\delta$ factor from pairing that changes the sign of the contribution  for odd-odd to even-even nuclei.  A combined fit to the data using Coulomb plus pairing contributions results in the dashed line in Fig.(\ref{fig:fig_12}), top panel.
The agreement with data is very good.   If we multiply the pairing energy 
by the factor $\delta$ we should get no discontinuities when plotting this quantity versus mass number.  Similarly if the properties of the free energy depend on the pairing
term, as for the ground state case, then it should be a monotonic function of $A$ after multiplying it by $\delta$.  In Fig.(\ref{fig:fig_12}) (bottom panel),  we plot the quantity $\frac{F}{T}\delta$ versus mass number for the same system of  Fig.(\ref{fig:fig_12}) (top panel).
The fit using the pairing mass dependence is also good. The Coulomb mass dependence fails especially for small mass number. From the values of the fit, using the ground state coefficients we can derive a temperature for the Coulomb case of $T=9.2 (\frac{\rho}{\rho_0})^{1/3}MeV$ where we have explicitly indicated
a possible density correction. For the pairing case we get $T=6.45MeV$.  Notice that in this case we have not suggested any density correction since 
the fate of the pairing energy at low density and finite temperature is `terra incognita'. When making a combined fit using pairing and Coulomb energy we get a good reproduction of the data (dashed line in Fig.(\ref{fig:fig_12}),top panel).
While the fitting value for pairing results in a `temperature' T=5.13 MeV, we get an increase of the Coulomb contribution to T=12.1 MeV.
Assuming that pairing is independent on density,
we could derive a density from the Coulomb result.  A simple calculation give $\frac{\rho}{\rho_0}=(6.45/9)^3=0.34$ which could be a reasonable indication of the density of the system when it breaks into fragments.

In summary in this section we have shown that the role of the Coulomb energy appears to be rather reduced in the reactions analyzed in this paper.  We expect it to become more
important for large nuclei.  On the other hand large nuclei have smaller symmetry and pairing energies per nucleon, thus a precise determination of the EOS can be obtained from measurements of isotopes having relatively
small masses.
\section{dynamics of the phase transition}
As we have seen  we have been able to discuss some observables in the fragmentation of nuclei using a language common to macroscopic systems undergoing a phase transition.  In the nuclear case we have 
a finite system composed at most of hundreds of particles which evolves in time under the influence of a  long range Coulomb force.  This poses many questions on why techniques of statistical mechanics should
apply in such  evolving nuclear systems.  This also offers the possibility of dealing with statistical mechanics of open systems and the problem of  extending the description  of a phase transition to such a system. 

We start by observing that even though we are dealing with a dynamical system, the order parameter defined in this work, $m$, is confined between -1 and +1.  In this sense we have a somewhat `closed' system.
Also the density at which the transition occurs should be smaller then normal density  and thus Coulomb effects are reduced.  However, if we deal with larger sources, such as in U+U collisions the phase transition might be washed out by the strong Coulomb field.
We expect our current considerations to be valid for small sources only.

   From statistical mechanics we know that in a first order phase transition~\cite{huang} a small seed increases in size depending upon the surface tension at a given T and density $\rho$.
If the pressure of the surrounding matter is smaller than the internal pressure of the drop, the drop will grow by capturing surrounding matter.  On the other hand if the opposite is true then the drop will decrease in size to balance the external pressure. The entire process is driven by surface tension.  Drops of a given size 
will survive only when their internal pressure balances the external pressure. If the system is at a very low density the interaction between different parts might take a relatively long time.
In these conditions a big nuclear drop whose internal pressure is larger than that of the surroundings could be considered to be  a nucleus which is evaporating particles in order to balance the external (zero in the case of an isolated nucleus) pressure.  If we accept this picture than the evaporation step is part of the dynamics of the phase transition.
Thus a very low density system might be thought of as  many isolated drops evaporating particles and reaching their equilibrium conditions before they collide with other parts of the system or as  small fragments being evaporated by other drops.  In a finite system
this does not happen, but we might think of a  process where at some point the finite system becomes unconfined  and an infinite system is approximated  by an infinite number of repetitions or `events'.  Of course in a statistically equilibrated system we know that time averages and event
averages are the same.  Here we are  extending  this concept to  finite systems where only event averages  can be used.  A  major question here is whether  the properties of the phase transition are decided very early, i.e. when the system `enters' the instability region.   As we said above if we have an infinite system at a very low density undergoing a first order phase transition, 
then the drops can explode, evaporate, and fuse with other particles over  a very long time.
Our finite system might behave similarly but without the fusion at later times.  If this were the case  than the detected fragments carry all the information of the phase transition, if not then we need to reconstruct the primary fragment distributions
coming out of the instability region.   
\begin{figure}[ht]
\centerline{
\includegraphics[width=8.3cm,clip]{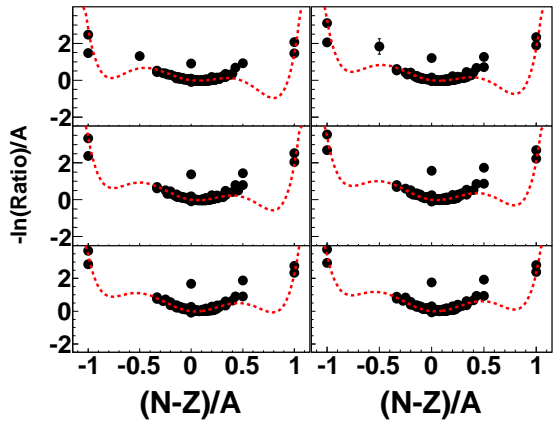}}
\caption{
Free energy vs.  time in AMD calculations  (see text) for $^{64}$Zn+$^{112}$Sn system at 40MeV/A and central collisions, i.e. impact parameter less than 3 fm. Different picture correspond to T=200, 300, 500, 1000, 1500 and 2000 fm/c respectively.}
\label{fig:fig_13}
\end{figure}

We can  try to clarify some of these questions by means of microscopic models such as Antisymmetrized Molecular Dynamics (AMD) or similar approaches  where the time evolution of
the system is followed~\cite{ono}.  However, we have to stress that in such microscopic models some assumptions are made  in order to recognize the fragments at particular times  during the time evolution. 
In simpler approaches, fragments are recognized if particles are close in coordinate space (of the order of the range of the attractive nuclear forces)~\cite{bon00}. In such a case the recognized fragments are 
`excited' and they evolve in time until a final state is reached after a long time of the order of thousands of fm/c.  A more refined approach for fragment recognitions is given by defining clusters when
its components are within a given distance in phase space.  The naive expectation would be that in this case we should recognize fragments earlier than the previous case and this is the method that we will adopt here
for simplicity following ref.~\cite{ono}.  In an ambitious approach~\cite{dorso} the claim is that fragments are recognized very early during the time evolution, of the order of tens of fm/c, if one searches for particles
connected in phase space to form fragments and minimize the energy.  This case probably corresponds to minimizing the entropy of the fragmenting system.  If this last picture will hold true then a picture of an infinite system at low density
will be equivalent to an `infinite' repetition of events.  Finally in all the considerations above we have to add the necessary and interesting complication that we have a mixture and not a single fluid, thus we can have more situations to explore 
than  discussed in the previous sections and we can `turn on and off' an external field as well. 
\begin{figure}[ht]
\centerline{
\includegraphics[width=8.3cm,clip]{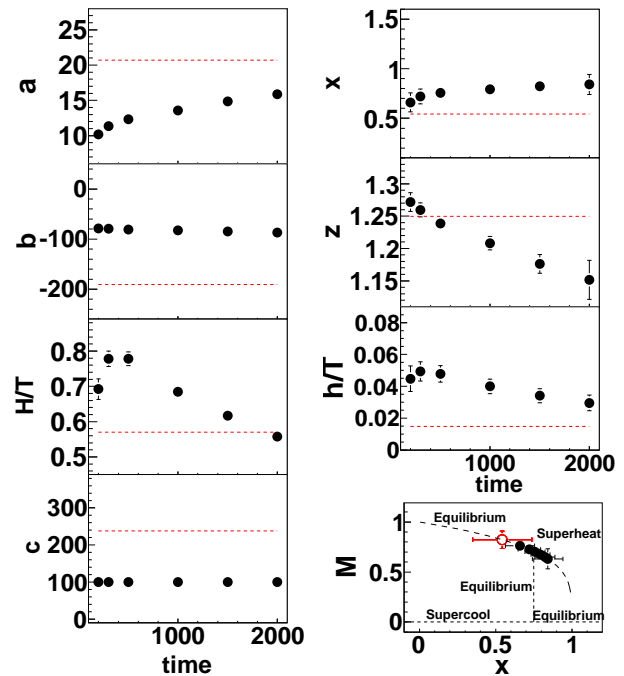}}
\caption{
Fit parameters a, b and c vs. time (see text) for the same system of Fig.\ref{fig:fig_13}. Solid circles refer to AMD calculations while the open symbol in the M vs. x plot is the experimental value for this system.}
\label{fig:fig_14}
\end{figure}

We have performed AMD calculations for the same systems investigated experimentally.  After some time t, fragments are separated enough in phase space that we can recognize within a simple
phase space coalescence approach as discussed in~\cite{ono}.  In this way we can define a yield at a given time and from this derive the free energy exactly as we did with the experimental data.
Characteristic results for the free energy versus time is given in Fig.(\ref{fig:fig_13}) together with a Landau $O(m^6)$ fit.  Some time evolution is  observed.  Using a more sophisticated fragment recognition approach~\cite{dorso} might even decrease the  time over which this evolution occurs.
We can study the time evolution  in more detail by plotting the variables a, b and c. M defined in the previous sections versus time.  The results of the fits to the free energy at different times is given in Fig.(\ref{fig:fig_14}).
While the quantities a, b and H/T change somewhat during the time evolution, smaller changes are observed in the time evolution of normalized quantities, x, z and h/T.  Nevertheless the time evolution of the fitting parameters
influences the time evolution of the order parameter M versus reduced temperature x as seen in the  bottom right of  Fig.(\ref{fig:fig_14}).  It is very interesting to see that in these units the system is initially very hot (superheated) and cools down when coming to  equilibrium below the critical temperature. The final result is very close to the observed values given by the open points.

Thus in this model most qualitative  features of the phase transition are decided very early during the time evolution. This might correspond to an entropy saturation early during the evolution.  However, different models and fragment recognition approaches might change the picture somewhat. 

\section{Equation of state}
\begin{figure}[ht]
\centerline{
\includegraphics[width=6.3cm,clip]{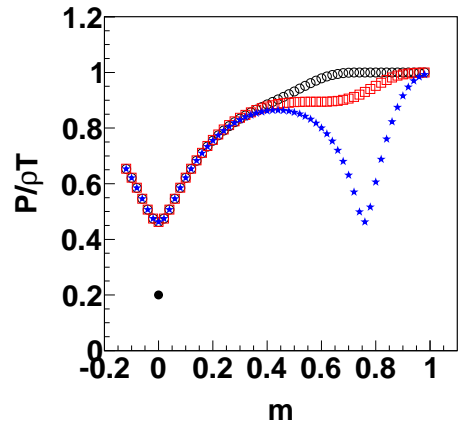}}
\caption{
Reduced pressure   versus m of the fragments obtained from the a, b and c parameters fit to the Landau free energy, Eq.(\ref{eq:eq_3}) for the $^{64}$Ni+$^{232}$Th.  The  curves correspond
to vapor(open circles), superheating(open squares), first order (3 critical line-solid stars), see text. The solid circle is for $N=Z$ nuclei at the critical point.}
\label{fig:fig_15}
\end{figure}

Once we know the free energy (at least in some cases) we can calculate the NEOS  by means of the Fisher model~\cite{fisher}.
Since we do not have at present experimental information on the density $\rho$, temperature T and pressure P of the system we can only estimate the
`reduced pressure'~\cite{paolo}:
\begin{equation}
  \label{eq:eq_18}
 \frac{P}{\rho T}(m)=\frac{M_0}{M_1 },
\end{equation}
where $M_i$ are moments of the mass distribution given by:
\begin{eqnarray}
\label{eq:eq_19}
M_k=\sum_A A^kY(A,m) =Y_0\sum_A A^k A^{-\tau}e^{-F/T(m) A}; \nonumber \\
k=0,...n.
\end{eqnarray}
 Notice that the quantities above are now dependent on the order parameter m.  From the knowledge of $F/T$ $(H/T=0)$ from the previous section we can easily
 calculate the reduced pressure near the critical point.  In particular given the simple expression for the moments we can also derive some analytical formulas following~\cite{paolo}:
 \begin{equation}
  \label{eq:eq_20}
 \frac{P}{\rho T}(m)=\frac{3.072|F/T|^{4/3}+1.417-3.631|F/T|+...}{-4.086|F/T|^{1/3}+3.631+0.966|F/T|+...},
\end{equation}
which gives at the critical point a critical compressibility factor ($F/T=0$):   $\frac{P}{\rho T}|_c=\frac{1.417}{3.631}=0.39$.\\
This value is essentially that derived from  the Van-der-Waals
 gas equation but is well above the values observed for real gases. Using the relations above we can calculate the NEOS for the situations illustrated in Fig.\ref{fig:fig_8}.  The results are displayed in Fig.\ref{fig:fig_15} where the reduced pressure is plotted
versus m for vaporization, superheating and first order phase transitions on the tri-critical line.  Notice that there is not a large difference between the first two cases, while
the last case displays two critical points (a third one is on the negative m axis).\\
  We have seen in Fig.\ref{fig:fig_1} that $N=Z$ nuclei display a power law.  We can also estimate the critical reduced 
pressure for this case noticing that the sums in Eq.(\ref{eq:eq_19}) above must be restricted to $A=2Z$ nuclei.  This leads to a critical compressibility factor  $\frac{P}{\rho T}_c=0.20$ which is a value closer to that estimated from other multi-fragmentation studies
before~\cite{mor}.\\

\begin{figure}[ht]
\centerline{
\includegraphics[width=6.3cm,clip]{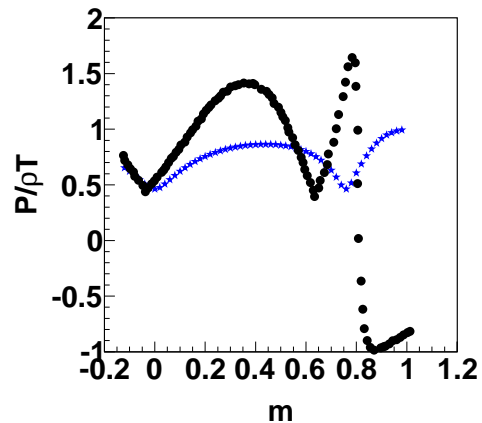}}
\caption{
Comparison to the analytical result, Eq.(\ref{eq:eq_20})(solid circles) for a first order phase transition.}
\label{fig:fig_16}
\end{figure}

We can compare our analytical result given in Eq.(\ref{eq:eq_20}) with the numerical values obtained above.  This is displayed in Fig.(\ref{fig:fig_16}) and we see that the numerical approximation is  
especially good near the critical point(s) as expected.  If, from detailed comparison to experimental data, we  are able to extract the temperature and pressure dependence of the parameters
entering the Landau free energy, then Eq.(\ref{eq:eq_18}) would be the Nuclear equation of state near a critical point.  From the actual data at our disposal we can only estimate the behavior of the reduced
pressure as function of the order parameter m.
On similar ground we can define a reduced compressibility as
 \begin{equation}
  \label{eq:eq_21}
  \chi  \rho T (m)=\frac{M_2}{M_1}.
\end{equation}
\begin{figure}[ht]
\centerline{
\includegraphics[width=6.3cm,clip]{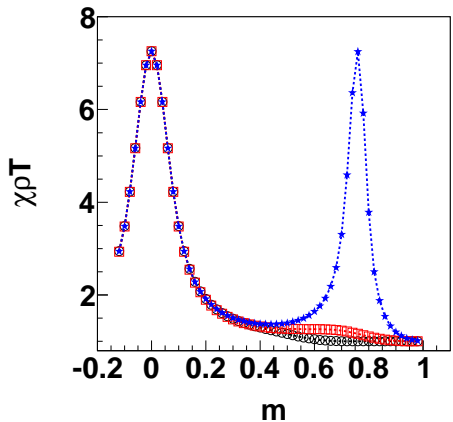}}
\caption{
Reduced compressibility   versus m of the fragments as in Fig.(\ref{fig:fig_15})}
\label{fig:fig_17}
\end{figure}
Its behavior is displayed in Fig.(\ref{fig:fig_17}) for the cases outlined above.  Divergences near the critical point(s) are obtained.

\section{conclusions}In conclusion, in this paper we have presented and discussed experimental 
evidence for the observation of a quantum phase transition in nuclei, 
driven by the neutron/proton asymmetry. Using the Landau approach, we 
have derived the free energies for our systems and found that they are 
consistent with the existence of a line of first-order phase transitions terminating at 
a point where the system undergoes a second-order transition. The 
properties of the critical point depend on the symmetry.
This is analogous to the well known superfluid  $\lambda$ transition 
in $^{3}$He-$^{4}$He mixtures. We suggest that a tricritical point, 
observed in $^{3}$He-$^{4}$He systems may also be observable in 
fragmenting nuclei.  These features call for further vigorous experimental 
investigation using high performance detector systems with excellent 
isotopic identification capabilities. Extension of these investigations 
to much larger asymmetries should be feasible as more exotic radioactive 
beams become available in the appropriate energy range.

It is important to stress that the observables discussed here represents only $necessary$ conditions for a critical behavior. A definite proof of a phase transition and a tricritical point could be given by a precise determination of
yields of fragments whose $m\approx \pm 0.5$, i.e. very unstable nuclei which, most probably, decay before reaching the detectors. Thus fragment-particle correlation  measurements for exotic primary fragments  such as $^4Li$, $^5Be$ (proton rich) or extremely neutron rich $^{10}He$ are needed. More generally, such correlation experiments can also shed light on the effects of secondary decay on the fragment observables. This remains a key question in many equation of state studies and model calculations differ in their assessment of these effects~\cite{wci,bon00}.   Higher quality data over a wider range of beam energies and colliding systems should also help in clarifying the role
of other energy terms, such as surface, Coulomb etc.,  which are important at lower excitation energies.  In particular the role of pairing and the possibility of  Bose-Einstein condensation,  should be more deeply investigated.  Our data for $I=0$ fragments already show that pairing is important. This might be due to its importance during the phase transition or to its role during secondary decay of the excited primary fragments. 
Exploration of quantum phase transitions in nuclei is important to our 
understanding of the nuclear equation of state and can have a significant 
impact in nuclear astrophysics, helping to clarify the evolution of 
massive stars, supernovae explosions and neutron star formation. \\

\begin{acknowledgments}
We thank the staff of the Texas A$\&$M Cyclotron facility for their support during the experiment. We thank L. Sobotka for letting us to use his spherical scattering chamber. This work is supported by the U.S. Department of Energy under Grant No. DE-FG03-93ER40773 and the Robert A. Welch Foundation under Grant A0330. One of us(Z. Chen) also thanks the \textquotedblleft100 Persons Project" of the Chinese Academy of Sciences for the support.
\end{acknowledgments}

\end{document}